\documentstyle[12pt]{article}

\def\hybrid{\topmargin -20pt	\oddsidemargin 0pt
	\headheight 0pt	\headsep 0pt
	\textwidth 6.25in	
	\textheight 9.5in	
	\marginparwidth .875in
	\parskip 5pt plus 1pt	\jot = 1.5ex}

\hybrid

\def\baselinestretch{1.2}

\catcode`\@=11

\def\marginnote#1{}
\newcount\hour
\newcount\minute
\newtoks\amorpm
\hour=\time\divide\hour by60
\minute=\time{\multiply\hour by60 \global\advance\minute by-\hour}
\edef\standardtime{{\ifnum\hour<12 \global\amorpm={am}%
	\else\global\amorpm={pm}\advance\hour by-12 \fi
	\ifnum\hour=0 \hour=12 \fi
	\number\hour:\ifnum\minute<10 0\fi\number\minute\the\amorpm}}
\edef\militarytime{\number\hour:\ifnum\minute<10 0\fi\number\minute}

\def\draftlabel#1{{\@bsphack\if@filesw {\let\thepage\relax
   \xdef\@gtempa{\write\@auxout{\string
      \newlabel{#1}{{\@currentlabel}{\thepage}}}}}\@gtempa
   \if@nobreak \ifvmode\nobreak\fi\fi\fi\@esphack}
	\gdef\@eqnlabel{#1}}
\def\@eqnlabel{}
\def\@vacuum{}
\def\draftmarginnote#1{\marginpar{\raggedright\scriptsize\tt#1}}

\def\draft{\oddsidemargin -.5truein
	\def\@oddfoot{\sl preliminary draft \hfil
	\rm\thepage\hfil\sl\today\quad\militarytime}
	\let\@evenfoot\@oddfoot	\overfullrule 3pt
	\let\label=\draftlabel
	\let\marginnote=\draftmarginnote
   \def\@eqnnum{(\theequation)\rlap{\kern\marginparsep\tt\@eqnlabel}%
\global\let\@eqnlabel\@vacuum}  }

\def\preprint{\twocolumn\sloppy\flushbottom\parindent 2em
	\leftmargini 2em\leftmarginv .5em\leftmarginvi .5em
	\oddsidemargin -.5in	\evensidemargin -.5in
	\columnsep .4in	\footheight 0pt
	\textwidth 10.in	\topmargin  -.4in
	\headheight 12pt \topskip .4in
	\textheight 6.9in \footskip 0pt
	\def\@oddhead{\thepage\hfil\addtocounter{page}{1}\thepage}
	\let\@evenhead\@oddhead	\def\@oddfoot{}	\def\@evenfoot{} }

\def\numberbysection{\@addtoreset{equation}{section}
	\def\theequation{\thesection.\arabic{equation}}}

\def\underline#1{\relax\ifmmode\@@underline#1\else
	$\@@underline{\hbox{#1}}$\relax\fi}

\def\titlepage{\@restonecolfalse\if@twocolumn\@restonecoltrue\onecolumn
     \else \newpage \fi \thispagestyle{empty}\c@page\z@
	\def\thefootnote{\fnsymbol{footnote}} }

\def\endtitlepage{\if@restonecol\twocolumn \else \newpage \fi
	\def\thefootnote{\arabic{footnote}}
	\setcounter{footnote}{0}}  

\catcode`@=12
\relax

\def\figcap{\section*{Figure Captions\markboth
	{FIGURECAPTIONS}{FIGURECAPTIONS}}\list
	{Figure \arabic{enumi}:\hfill}{\settowidth\labelwidth{Figure
999:}
	\leftmargin\labelwidth
	\advance\leftmargin\labelsep\usecounter{enumi}}}
 \relax
\def\tablecap{\section*{Table Captions\markboth
	{TABLECAPTIONS}{TABLECAPTIONS}}\list
	{Table \arabic{enumi}:\hfill}{\settowidth\labelwidth{Table
999:}
	\leftmargin\labelwidth
	\advance\leftmargin\labelsep\usecounter{enumi}}}
 \relax
\def\reflist{\section*{References\markboth
	{REFLIST}{REFLIST}}\list
	{[\arabic{enumi}]\hfill}{\settowidth\labelwidth{[999]}
	\leftmargin\labelwidth
	\advance\leftmargin\labelsep\usecounter{enumi}}}
 \relax
\makeatletter
\newcounter{pubctr}
\def\publist{\@ifnextchar[{\@publist}{\@@publist}}
\def\@publist[#1]{\list
	{[\arabic{pubctr}]\hfill}{\settowidth\labelwidth{[999]}
	\leftmargin\labelwidth
	\advance\leftmargin\labelsep
	\@nmbrlisttrue\def\@listctr{pubctr}
	\setcounter{pubctr}{#1}\addtocounter{pubctr}{-1}}}
\def\@@publist{\list
	{[\arabic{pubctr}]\hfill}{\settowidth\labelwidth{[999]}
	\leftmargin\labelwidth
	\advance\leftmargin\labelsep
	\@nmbrlisttrue\def\@listctr{pubctr}}}
 \relax
\makeatother
\newskip\humongous \humongous=0pt plus 1000pt minus 1000pt

\newif\ifdtup

\relax

\def\thefootnote{\fnsymbol{footnote}}
\def\be{\begin{equation}}
\def\ee{\end{equation}}
\def\ba{\begin{eqnarray}}
\def\ea{\end{eqnarray}}

\begin{document}
\renewcommand{\theequation}{\arabic{equation}}
\newcommand{\beq}{\begin{equation}}
\newcommand{\eeq}[1]{\label{#1}\end{equation}}
\newcommand{\ber}{\begin{eqnarray}}
\newcommand{\eer}[1]{\label{#1}\end{eqnarray}}
\begin{titlepage}
\begin{center}

\hfill CERN--TH.7499/94\\
\hfill hep-th/9411118\\

\vskip .5in

{\large \bf STRING DUALITIES AND THE GEROCH GROUP}

\vskip 0.5in

{\bf Ioannis Bakas}
\footnote{Permanent address: Department of Physics, University of
Crete,
GR--71409 Heraklion, Greece}
\footnote{e--mail address: BAKAS@SURYA11.CERN.CH}\\
\vskip .1in

{\em Theory Division, CERN\\
     CH-1211 Geneva 23, Switzerland}\\

\vskip .1in

\end{center}

\vskip 1.1in

\begin{center} {\bf ABSTRACT } \end{center}
\begin{quotation}\noindent
We examine the properties and symmetries of the lowest order
effective theory of 4--dim string backgrounds with
axion and dilaton fields and zero cosmological constant.
The dimensional reduction yields an
$\hat{O}(2,2)$ current group of transformations in the
presence of two commuting Killing symmetries. Special emphasis
is placed on the identification of the T and S string duality
symmetries, and their intertwining relations.

\vskip2.8cm
\noindent
{\em Contributed to the Proceedings of the Satellite Colloquium
``Topology, Strings and Integrable Models" to the XIth
International Congress of Mathematical Physics, Paris, 25--28
July 1994; Diderot Editeur.}

\end{quotation}
\vskip2.0cm
CERN--TH.7499/94\\
November 1994\\
\end{titlepage}
\vfill
\eject

\def\baselinestretch{1.2}
\baselineskip 16 pt
\noindent
The string dualities are symmetries beyond the ordinary
reparametrization
invariance of general relativity, which provide a novel view of space
time,
as experienced by extended objects propagating in it in a quantum
mechanically consistent way. As such, they are considered to be
instrumental in the understanding of various aspects of the string
equivalence
principle, a concept whose general formulation is admittedly lacking
at
the moment. T--duality, which is the best known example of this kind,
is
associated with string backgrounds that possess an Abelian isometry.
S--duality, which has received
attention more recently in the context of the toroidal
compactification
of the heterotic string, is a discrete $SL(2,Z)$ transformation that
acts
on the coupling constant of the theory  and has a non--perturbative
meaning in the full theory.

It is natural to expect that the string dualities will be discrete
remnants
of continuous symmetries of the lowest order effective field theory
that
models the vanishing conditions of all beta functions in the 2--dim
$\sigma$--model approach to string dynamics. This is certainly true
for
the known dualities, which involve discrete transformations of the
massless modes of the string, namely the target space metric $G_{\mu
\nu}$,
the antisymmetric tensor field $B_{\mu \nu}$ and the dilaton $\Phi$.
So, apart from using the continuous symmetries of the effective
theory as
solution generating techniques for non--trivial string backgrounds to
lowest order in ${\alpha}^{\prime}$, there is also the scope for
discovering new duality transformations within that framework. For
example,
if a class of string models exhibit both T and S--duality symmetries,
their
intertwining will give rise to new symmetries, like T--S--T, etc. It
is
important to emphasize that the solution generating transformations
of ordinary gravitational theories have nothing to do with general
covariance, once the diffeomorphism group is moded out. But if the
diffeomorphism group is moded out in string theory, there will still
remain
discrete duality symmetries attributed to the string equivalence
principle.
We will study the properties and symmetries of the string effective
theory
with this particular philosophy in mind.

We consider strings in $M_{4} \times K$, where $M_{4}$ is a 4--dim
space with signature $-+++$ and $K$ is some internal space whose
details
will be irrelevant for the present purposes. We assume that $c(M_{4})
= 4$,
i.e. the cosmological constant is zero (at least to lower order in
${\alpha}^{\prime}$) and perform the dimensional reduction of the
string
background equations in the presence of two Abelian isometries in
$M_{4}$.
We make the ansatz for the metric $G_{\mu \nu}$,
\be
ds^2 = f(Z_{+}, Z_{-})dZ_{+}dZ_{-} + g_{AB}(Z_{+}, Z_{-})
dX^{A}dX^{B}
{}~ ; ~~~~ A, ~ B = 2, ~ 3 ~ ,
\ee
and we also take $\Phi(Z_{\pm})$ and $b(Z_{\pm})$, where $b$ is the
axion
field associated to $B_{\mu \nu}$. It then follows that the only
non--trivial component of the antisymmetric tensor field is
$B_{23} \equiv B$ and that
\be
{\partial}_{\pm}b = \pm {e^{-4 \Phi} \over \sqrt{\det g}}
{\partial}_{\pm} B ~ .
\ee
We are considering this special class of 4--dim string backgrounds
because
the dimensionally reduced theory is quite rich and exhibits an
infinite
dimensional hidden symmetry that acts on the space of solutions. In
fact, the
reduced theory is an integrable 2--dim system consisting of two
(essentially)
decoupled Ernst $SL(2,R)/U(1)$ $\sigma$--models, one for the
symmetric
matrix $g$ and the other for the axion--dilaton matrix
\be
\lambda = \sqrt{\det g} ~ e^{2 \Phi} \left(\begin{array}{ccc}
1 &  & b \\
b &  & b^2 + e^{-4 \Phi} \end{array} \right) ~ .
\ee
The symmetry that transforms backgrounds of this type into each other
is
a string generalization of the Geroch group that is encountered in
reduced vacuum gravity.

The Ernst $\sigma$--model equation that encodes all non--linearities
of
the reduced string background equations is
\be
{\partial}_{+}(\sqrt{\det g} ~ g^{-1}{\partial}_{-}g) +
{\partial}_{-}(\sqrt{\det g} ~ g^{-1}{\partial}_{+}g) = 0 ~.
\ee
We also have the same equation for $\lambda$, while the
conformal factor $f(Z_{\pm})$ satisfies linear first order
differential equations that can be integrated by quadratures,
once $g$ and $\lambda$
have been determined. The symmetries of the Ernst model can be
described using the twist potential $\psi(Z_{\pm})$ of $g$,
${\partial}_{\pm} \psi = \pm \epsilon \sqrt{\det g} ~
g^{-1}{\partial}_{\pm}g$,
where $\epsilon$ is the constant $2 \times 2$ antisymmetric matrix.
Then, it can be verified that the variations
\be
{\delta}_{T}^{(0)}g = [g \epsilon , ~ T] \epsilon ~ , ~~~~
{\delta}_{T}^{(1)}g = [g \epsilon , ~ T \psi \epsilon -
2 \psi \epsilon T + {\psi}^{t} \epsilon T] \epsilon
\ee
are infinitesimal symmetries of the 2--dim model,
provided that $T$ is a generic
$SL(2, R)$ Lie algebra element. By intertwining these transformations
we
obtain an infinite dimensional symmetry algebra,
$[{\delta}_{T}^{(n)} , ~ {\delta}_{T^{\prime}}^{(m)}] =
{\delta}_{[T, ~ T^{\prime}]}^{(n+m)}$, which is an $\widehat{SL}
(2,R)$ current
algebra known as the Geroch symmetry. Repeating the same analysis for
the
$\lambda$ matrix and taking into account the fact the
${\delta}_{T}^{(n)} \sqrt{\det g} = 0$, we conclude that the hidden
symmetry of the 4--dim string backgrounds in question is an
$\hat{O} (2,2)$ current group. This symmetry, which is conveniently
formulated in the Einstein frame, is clearly non--locally
realized for all $n > 0$.

The string Geroch group incorporates both T and S--duality
symmetries.
Explicit calculation shows that in the presence of two Abelian
isometries, the continuous $O(2,2)$ transformations underlying
T--duality have an $SL(2)$ factor generated by the zero modes of
the $g$--$\widehat{SL} (2)$ Geroch symmetry, while the other $SL(2)$
factor is generated by the modes ${\delta}_{-}^{(1)}$,
${\delta}_{0}^{(0)}$, ${\delta}_{+}^{(-1)}$ of the
$\lambda$--$\widehat{SL} (2)$ Geroch symmetry. As for the continuous
$SL(2)$
transformations underlying S--duality, its generators can be
identified
with the zero modes of the $\lambda$--$\widehat{SL} (2)$ Geroch
symmetry. Hence, apart from providing a unifying framework for both
T and S--duality symmetries, we also observe that the intertwining of
T and S produces infinitely many more additional symmetries, which
are
associated to higher modes of the string Geroch group. It is
reasonable to
expect that the complete group of dualities for this particular
sector
of string theory will be an infinite dimensional discrete subgroup of
$\hat{O} (2,2)$ realized non--locally in the Einstein (and in the
$\sigma$--model) frame of the theory. Their world--sheet origin is
lacking at the moment.

More details can be found in the work {\em ``O(2,2)
Transformations and the String Geroch Group"} by the
present author, published in Nucl. Phys.
\underline{B428} (1994) 374 (and references therein).

\end{document}